\documentclass[aps,prl,reprint,twocolumn,amsmath,amssymb]{revtex4}
\usepackage{epsfig,amssymb}
\usepackage{graphicx}
\usepackage{dcolumn}
\usepackage{bm}
\usepackage{dcolumn}
\usepackage{multirow}
\begin{document}
\title{First-principles calculations of the electronic structure of open-shell condensed matter systems}
\author{Johannes~Lischner, Jack~Deslippe, Manish~Jain, Steven~G.~Louie}
\affiliation{Department of Physics, University of California, Berkeley, California 94720, USA, and Materials Sciences Division, Lawrence Berkeley National Laboratory, Berkeley 94720, USA.}

\begin{abstract}
We develop a Green's function approach to quasiparticle excitations of
open-shell systems within the GW approximation. It is shown that
accurate calculations of the characteristic multiplet structure
require a precise knowledge of the self energy and, in particular, its
poles. We achieve this by constructing the self energy from
appropriately chosen mean-field theories on a fine frequency grid. We
apply our method to a two-site Hubbard model, several molecules and
the negatively charged nitrogen-vacancy defect in diamond, and obtain
good agreement with experiment and other high-level theories.

\end{abstract}

\pacs{71.15.Mb}
\maketitle
\emph{Introduction}.---In nature, there exists a wide range of
electronic systems with open shells, including most atoms and many
molecules, but also defects in crystalline solids. These systems play
important roles in almost all areas of physics, chemistry and biology:
for example, the negatively charged nitrogen-vacancy (NV$^{-}$)
defects in diamond are used for biological imaging \cite{Lukin,Fann}
and are also promising candidates for qubits in quantum computers
\cite{Wrachtrup,Hanson,Awschalom}.

It is therefore important to develop theoretical methods to study
open-shell systems and their properties. While for closed-shell
systems a well-established set of methods exists, ranging from wave
function-based quantum chemistry approaches to density-functional
theory (DFT) and Green's function based many-body perturbation theory,
the accuracy of these methods when applied to open-shell systems is
less certain: Even the application of wave function-based methods to
small open-shell molecules is far from straightforward \cite{Stanton}
and standard density functionals are known to break the orbital and
spin degeneracy of the ground state \cite{Savin,FertigKohn}.

A Green's function approach to electron excitations in open-shell
system was first considered by Cederbaum and coworkers
\cite{CederbaumDomcke,CederbaumSchirmer} in the 1970's. These authors
only applied the formalism to toy problems with few orbitals and
employed approximations to the self energy which are not feasible for
large systems, such as open-shell defects. Other previous applications
of Green's function theory to open-shell systems using the GW
approximation had either carefully selected reference states to avoid
complications associated with the open-shell \cite{ShirleyMartin} or
ignored the degenerate ground-state problem \cite{MaRohlfing,Rubio}.

In this Letter, we extend the GW approach to open-shell
systems. Calculations on several prototypical systems are performed: a
two-site Hubbard cluster, four molecules (nitrogen dioxide, oxygen,
nitrogen difluoride, chlorine dioxide) and the NV$^{-}$ center in
diamond.  We find our approach is capable of describing these systems
with quantitative accuracy. We have identified and implemented two
important elements for accurate results in GW calculations of
open-shell systems: i) a careful choice of the mean-field starting
point providing accurate self-energy pole positions, and ii) a method
for evaluating the self energy on a fine frequency grid.

\emph{Theory}.---In a photoemission experiment with photons of energy
$\omega_{\text{photon}}$ (setting $\hbar=1$), the photocurrent
$J(\epsilon_{\bm{k}})$ due to photoelectrons with momentum $\bm{k}$
and energy $\epsilon_{\bm{k}}$ is given by \cite{Hedin}
\begin{align}
J(\epsilon_{\bm{k}}) = \sum_{ij} \Delta_{\bm{k}i}\Delta_{j\bm{k}}
A_{ij}(\epsilon_{\bm{k}}-\omega_{\text{photon}}),
\end{align}
where $\Delta_{\bm{k}i}=\langle
\bm{k}|\Delta_{\text{dipole}}|\psi_i\rangle$ and $A_{ij}(\omega)=
\langle \psi_i|A(\bm{r},\bm{r}',\omega)|\psi_j \rangle$ denote matrix
elements of the dipole operator and the spectral function,
respectively, with $\psi_i$ being an appropriate single-particle
orbital. Neglecting off-diagonal matrix elements for an appropriately
chosen physical set of orbitals, we obtain $A_{jj}(\omega)=1/\pi
|\text{Im} G_{jj}(\omega)|$ by computing the interacting Green's
function (here we give the electron removal part)
\begin{align}
G_{jj}(\omega) = \sum_\lambda \frac{|\langle N-1,\lambda|c_j|N,0\rangle|^2}
{\omega - E_\lambda - i\eta}
\label{eq:lehmann}
\end{align}
with $E_\lambda = E^{(N)}_0 - E^{(N-1)}_\lambda$. Here, $|N,0\rangle$
and $E^{(N)}_{0}$ denote the $N$-particle ground state and its energy,
respectively, while $|N-1,\lambda \rangle$ denotes an $(N-1)$-particle
state (with $\lambda$ being an appropriate set of quantum numbers) with
energy $E^{(N-1)}_\lambda$. Also, $c_j$ is the destruction operator
for an electron in orbital $j$ and $\eta=0^+$.

$E_\lambda$ solves the quasiparticle equation
\begin{equation}
E_{\lambda} =\epsilon_{j}+\Sigma_{jj}(E_{\lambda})-V^{xc}_{jj},
\label{eq:qp}
\end{equation}
where $\epsilon_{j}$ and $V^{xc}_{jj}$ denote the orbital energy and a
diagonal matrix element of the exchange-correlation potential from a
mean-field calculation, respectively, while $\Sigma_{jj}(\omega)$ is a
diagonal matrix element of the self-energy operator.

The quasiparticle equation [Eq.~\eqref{eq:qp}] follows from Dyson's
equation \cite{FetterWalecka}
\begin{equation}
G^{-1}_{ij}(\omega)=G^{-1}_{0,ij}(\omega) - \Sigma_{ij}(\omega) + V^{xc}_{ij},
\label{eq:dyson}
\end{equation}
which relates the interacting Green's function to the mean-field
Green's function $G_{0,ij}(\omega)$ via the self energy. The standard
derivation of Dyson's equation \cite{FetterWalecka} assumes the
existence a nondegenerate interacting ground state which evolves into
a nondegenerate single Slater determinant state as the interactions
are adiabatically turned off. The hallmark of open-shell systems,
however, is the existence of multiple degenerate ground states which
do not generally evolve into noninteracting single Slater determinant
states \cite{Brouder}. If --- for a particular ground state --- the
resulting noninteracting state is a sum of Slater determinants, one
has to employ the methods of quantum field theory with initial
correlations and replace Dyson's equation with a more complicated
expression \cite{Brouder,Hall}. In our calculations, we avoid this
difficulty by carefully choosing a ground state which evolves into a
single Slater determinant such that Dyson's equation is valid. In
particular, we work with the ground state with the highest magnetic
quantum number because there exists a corresponding single Slater
determinant with the same properties (i.e., it is also an eigenstate
of the total spin and/or orbital angular momentum operator with the
same eigenvalue) \cite{CederbaumSchirmer}. An approximation to this
particular ground state is provided by standard spin-polarized
mean-field calculations. We note that it is not always possible to
find a single determinant ground state. However, such a state must
exist whenever Hund's rules apply.

In closed-shell systems, Eq.~\eqref{eq:qp} typically has a single
solution leading to a pronounced quasiparticle peak in
$A_{jj}(\omega)$ which corresponds to the removal of an electron from
orbital $j$ \cite{LouieHybertsen}. In open-shell systems, the orbital
and spin angular momenta of the electrons in the unfilled shells can
couple in various ways resulting in \emph{multiple} low-energy
eigenstates of the $N$ and the $(N-1)$-particle system. The coupling
of angular momenta generally produces eigenstates which are sums of
\emph{multiple} Slater determinants \cite{CohenTannoudji}. As a
consequence, \emph{multiple} eigenstates of the $(N-1)$-particle
system can make significant contributions to $G_{jj}(\omega)$ if their
matrix element in the numerator of Eq.~\eqref{eq:lehmann} is
large. $G_{jj}(\omega)$ then has \emph{multiple} poles and we expect
to find \emph{multiple} solutions of Eq.~\eqref{eq:qp}. This important
connection between the poles of the self energy and the multiplet
structure of open-shell systems was first established by Cederbaum and
coworkers \cite{CederbaumDomcke,CederbaumSchirmer}.

If $G_{jj}(\omega)$ has multiple poles, Eq.~\eqref{eq:dyson} shows
that the self energy $\Sigma_{jj}(\omega)$ must also have poles
occurring \emph{between} the poles of $G_{jj}(\omega)$. The occurrence
of poles in $\Sigma_{jj}(\omega)$ near $E_\lambda$ is a particular
feature of open-shell systems and a direct consequence of the
electronic multiplet structure.

In actual calculations for open-shell systems, a precise knowledge of
the \emph{frequency dependence} of the self energy is necessary to
locate its poles and obtain accurate multiplet splittings. In
contrast, for closed-shell systems it is usually sufficient to employ
a simple linear expression for the frequency dependence of the self
energy in the vicinity of the quasiparticle energy
\cite{LouieHybertsen}.

In this work, we employ the GW approximation to the self energy
following the first-principles method of Hybertsen and Louie
\cite{LouieHybertsen}.  To obtain $\Sigma_{jj}(\omega)$ at many
frequencies, we make use of a specific form of the evaluation of the
frequency dependence of the dielectric response and self energy as
proposed in Refs. \cite{Tiago} and \cite{ShirleyMartin}.  In this
approach, $\Sigma_{jj}(\omega)$ is separated into a
frequency-independent bare exchange part $\Sigma^{(x)}_{jj}$ and a
frequency-dependent correlation part $\Sigma^{(c)}_{jj}(\omega)$ given
by
\begin{equation}
\Sigma^{(c)}_{jj}(\omega) = \sum_{nI} 
\frac{|V_{jnI}|^2}{\omega - \epsilon_{n}-\Omega_I sgn(\epsilon_{n}-\mu)},
\label{eq:sigmaGW}
\end{equation}
where $\mu$ denotes the chemical potential and $\Omega_I$ is a neutral
excitation energy of the $N$-particle system obtained by solving
Casida's equation in the random-phase approximation
\cite{Tiago}. Also, $V_{jnI}$ denotes a Coulomb matrix element between
the product $\psi^*_j\psi_n$ and the fluctuation charge density
$\rho_I$ \cite{Tiago} (see Supplementary Material for details on the
approach).

Equation~\eqref{eq:sigmaGW} shows that the poles of
$\Sigma_{jj}(\omega)$ are determined by the mean-field electron
removal (or addition) energies $\epsilon_{n}$, which are the poles of
$G_{0}$, and by the neutral excitation energies $\Omega_I$, which are
the poles of the screened interaction $W_0$ in the random-phase
approximation. Both $\epsilon_n$ and $\Omega_I$ depend on the
mean-field theory used to compute $G_0$ and $W_0$, implying an
analogous dependence on the choice of the mean-field starting point
for the poles of $\Sigma_{jj}(\omega)$.

In principle, the self energy should be computed from the interacting
Green's function $G$, whose poles are at $E_\lambda$, and the exact
screened interaction $W$ \cite{LouieHybertsen}. For closed-shell
systems, it is possible to carry out self-consistent GW$_0$
calculations where the self energy is recomputed using the iterated
Green's functions such that $\Sigma$ becomes independent of the
mean-field starting point \cite{ShishkinKresse}. For open-shell
systems, self-consistent calculations are more difficult because of
the more complicated structure of $G$ and additional problems to be
discussed below. To obtain accurate self-energy pole positions we
instead carefully choose mean-field theories that yield $\epsilon_n$
and $\Omega_I$ which are good approximations to $E_\lambda$ and the
poles of the exact $W$, respectively. In general, one finds that the
poles of $W_0$ obtained from standard density-functional calculations
are good approximations to neutral excitation energies. In contrast,
the poles of $G_0$ obtained from density-functional theory often
differ from the exact removal or addition energies (i.e. the
quasiparticle energies) by several electron volts. Such an error in
the poles of $G_0$ leads to a similar-sized error in the self energy
pole locations and to a large error in the multiplet splittings. To
obtain the best $G_0$, we construct it from mean-field calculations
using the static COHSEX approximation \cite{DeslippeCOHSEX,Reining}.

In addition, if the result of a calculation depends on a particular
self-energy pole we carry out partially self-consistent calculations
where we only update the particular $\epsilon_n$ in
Eq.~\eqref{eq:sigmaGW} which determines the position of the self-energy
pole under consideration.

\emph{Molecules}.--- First, we study the electronic multiplet
structure of four small molecules for which accurate experimental data
is available.

Nitrogen dixoide (NO$_2$) has a doublet ground state. We first carry
out DFT calculations \cite{KohnSham,KohnHohenberg} at the experimental
geometry \cite{Edquist} using the spin-polarized LDA
exchange-correlation functional, norm-conserving pseudopotentials, a
plane-wave basis (50 Ry cutoff) and a cubic supercell with linear
dimension of $10.6~\AA$.

For the construction of $W_0$ we use wave functions and energies from
the DFT calculation. We use 300 empty states and a 15 Ry momentum
space cutoff for the dielectric response. For $G_0$ we use wave
functions and energies from a static COHSEX
calculation. Table~\ref{NO2dat} shows that the COHSEX single-particle
energies are much closer to the experimental ionization potentials
than the DFT energies, but the multiplet structure is still missing in
this calculation. For the calculation of the self-energy matrix
element we use 300 empty states and a modified static remainder
correction \cite{DeslippeREMAINDER,Hybertsen} which extends the sum
over $n$ in Eq.~\eqref{eq:sigmaGW} to \emph{all} empty states and
greatly improves convergence. This choice of parameters results in
multiplet splittings converged to within $\sim 0.1$~eV.

\begin{figure}
  \includegraphics[width=8.cm]{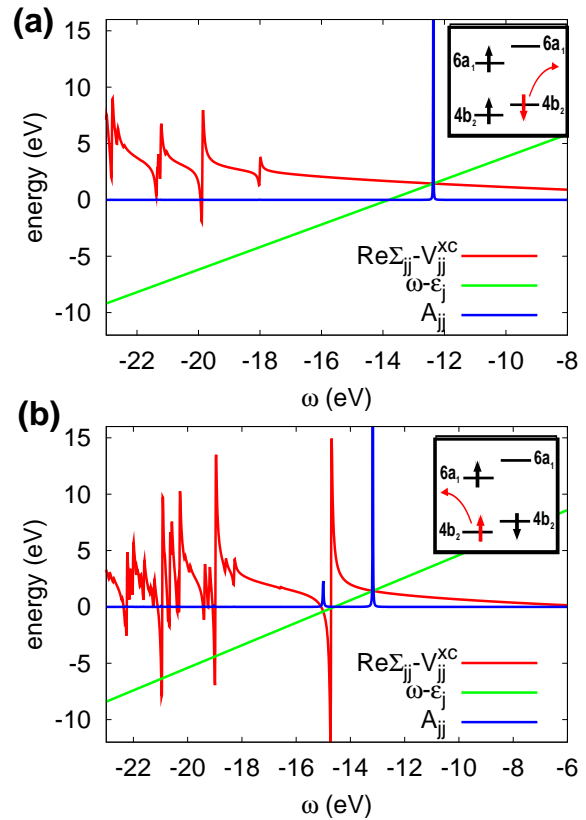} 
  \caption{Self energy $\Sigma_{jj}(\omega)$ and spectral function
    $A_{jj}(\omega)$ for (a) the removal of a down-spin electron from
    the $j=4b_2$ orbital in NO$_2$ and (b) the removal of an up-spin
    electron from the $j=4b_2$ orbital. A Lorentzian
    broadening of 20~meV is used for each curve.}
  \label{fig:8D}
\end{figure}

Figure~\ref{fig:8D}(a) shows the self energy and spectral function for
the removal of a \emph{down-spin} electron from the $4b_2$ orbital
[see insert in Fig.~\ref{fig:8D}(a)]. We do not expect any multiplet
structure for this process because the up-spin hole can only couple to
the up-spin electron in the $6a_1$ orbital to give a triplet
state. Indeed, the spectral function exhibits a single peak
corresponding to the triplet ($^3B_2$) state.

Figure~\ref{fig:8D}(b) shows results for the removal of an
\emph{up-spin} electron from the $4b_2$ orbital. The down-spin hole
can now couple to the up-spin electron in the $6a_1$ orbital to yield
either a singlet ($^1B_2$) or a triplet ($^3B_2$) state. Indeed, we
find two solutions of Eq.~\eqref{eq:qp} resulting in two poles of the
Green's function and two peaks in the spectral function with a
singlet-triplet splitting of 1.8~eV which compares favorably with the
experimental splitting of 1.5~eV (Table~\ref{NO2dat}). In contrast,
the singlet-triplet splitting from G$_{LDA}$W$_{LDA}$ is 2.9~eV
highlighting the importance of an accurate mean-field starting
point. To make sure that the two solutions are indeed multiplet states
we traced back the low lying self-energy pole to open-shell features
in G$_0$ and W$_0$: namely, to the pole in G$_0$ due to the unpaired
up-spin 6a$_1$ state and the pole in W$_0$ due to the
4b$_{2\downarrow}$~$\rightarrow$ 6b$_{1\downarrow}$ transition between
the two open shells. Hund's rule suggests that the lower energy
solution is the triplet state.

\begin{table}
  \setlength{\doublerulesep}{1\doublerulesep}
  \setlength{\tabcolsep}{2\tabcolsep}
    \caption{Comparison of our results for NO$_2$ with experiment
      \cite{Edquist}. All energies are given in eV.}
  \begin{ruledtabular}
    
    \begin{tabular}{c c c c c c c}
      orbital & DFT & COHSEX & G$_{LDA}$W$_{LDA}$ & GW & exp. & state \\
      \hline
      6a$_1$($\uparrow$)   & -6.6 & -12.0 & -10.7 & -11.2  & -11.2 & $^1A_1$  \\
      4b$_2$($\downarrow$) & -8.7 & -14.1 & -12.5 & -12.8  & -13.0 & $^3B_2$ \\
      4b$_2$($\uparrow$)   & -9.3 & -14.9 & -10.5 & -13.6  & -13.0 & $^3B_2$ \\
      4b$_2$($\uparrow$)   & -9.3 & -14.9 & -13.4 & -15.4  & -14.5 & $^1B_2$ 
    \end{tabular}

  \end{ruledtabular}
  \label{NO2dat}
\end{table}

Inspection of Table~\ref{NO2dat} shows that we obtain \emph{two}
values for the energy of the triplet state $^3B_2$, one from the
removal of an \emph{up-spin} electron from the $4b_2$ orbital, one
from the removal of a \emph{down-spin} electron from the same
orbital. These values differ by $0.8$~eV and bracket the experimental
result. There are two factors which contribute to this discrepancy: i)
remaining errors in the positions of the self-energy poles which
contaminate only solutions of the up-spin quasiparticle equation and
ii) missing vertex corrections which contaminate solutions of the up-
and down-spin quasiparticle equations in different amounts
\cite{ShirleyThesis}. We expect that the inclusion of vertex
corrections will reduce the difference. Nevertheless, as shown above,
accurate multiplet splittings can be extracted from our calculations
if the energy differences are calculated from solutions of the
quasiparticle equation for a \emph{particular spin direction}.

The ratio of the areas under the singlet and the triplet peaks in
Fig.~\ref{fig:8D}(b) should be the experimentally observed ratio of
photoemission intensities, the so-called multiplet ratio
\cite{CederbaumMultiplet}. We find in our calculations that the
multiplet ratios are much more sensitive to the positions of the
self-energy poles than the multiplet splittings. We do not expect that
these ratios can be computed reliably with our current GW approach
because of the remaining uncertainties in the self-energy pole
locations.  However, Schirmer and coworkers found a relatively simple
analytical procedure for calculating these ratios based on the
addition of angular momenta \cite{CederbaumMultiplet}. We expect that
the combination of their approach for the multiplet ratios and the GW
approach for the multiplet splittings offers a reliable and
\emph{complete} description of the multiplet structure of open-shell
systems.

Table~\ref{MolDat} shows our results for the oxygen (O$_2$), nitrogen
difluoride (NF$_2$) and the chlorine dioxide (ClO$_2$) molecules. The
GW multiplet splittings are 2.4~eV for O$_2$, 2.3~eV for NF$_2$ and
2.5~eV for ClO$_2$. They compare favorably with experimental
splittings: 2.3~eV for O$_2$, 1.8~eV for NF$_2$ and 2.4~eV for ClO$_2$
\cite{O2exp,NF2exp,ClO2exp}. However, splittings obtained from
G$_{LDA}$W$_{LDA}$ can deviate from experimental findings by several
electron volts.

\begin{table}
  \setlength{\doublerulesep}{1\doublerulesep}
  \setlength{\tabcolsep}{2\tabcolsep}
  \caption{Comparison of our results for O$_2$, NF$_2$ and ClO$_2$
    with experiment \cite{O2exp,NF2exp,ClO2exp}. All energies are given in eV.}
  \begin{ruledtabular}
    
    \begin{tabular}{c c c c c c}
         & orbital & state &     G$_{LDA}$W$_{LDA}$& GW & exp.  \\ 
      \hline
      O$_2$ & 1$\pi_g$ &   $^2\Pi_g$ &      12.1 & 12.4 & 12.3 \\ 
      O$_2$ & 3$\sigma_g$& $^4\Sigma^-_g$ & 15.4 & 19.2 & 18.4 \\ 
      O$_2$ & 3$\sigma_g$& $^2\Sigma^-_g$ & 19.5 & 21.6 & 20.7 \\
      \hline
      NF$_2$& 2b$_1$ &     $^1$A$_1$&       11.6 & 12.0 & 12.1 \\ 
      NF$_2$& 6a$_1$ &     $^3$B$_1$&       11.9 & 15.0 & 14.6 \\ 
      NF$_2$& 6a$_1$ &     $^1$B$_1$&       14.7 & 17.3 & 16.4 \\ 
      \hline
      ClO$_2$& 3b$_1$ &    $^1$A$_1$&       10.2 & 10.6 & 10.5 \\ 
      ClO$_2$& 1a$_2$ &    $^3$B$_1$&       10.5 & 13.3 & 13.0 \\ 
      ClO$_2$& 1a$_2$ &    $^1$B$_1$&       13.5 & 15.8 & 15.4 \\
    \end{tabular}

  \end{ruledtabular}
  \label{MolDat}
\end{table}

\emph{Hubbard cluster}.--- To further establish the accuracy of our
method, we apply it to an analytically solvable model system: a
two-site Hubbard cluster containing three electrons with a Hilbert
space spanned by four spin-orbitals. We denote the hopping parameter
$t$ and the on-site interaction $U$. This system has a doublet ground
state. We compute the mean-field wavefunctions and energies using the
spin-polarized Hartree-Fock method and then evaluate the self energy
corresponding to the removal of an up-spin electron from the doubly
occupied bonding orbital. We find two solutions of the quasiparticle
equation due to the occurence of a pole in the self energy: their
separation is $1.1 U$ for $U/t < 1$ where we expect the GW
approximation to give accurate results.

In this model system, analytical evaluation of the Hamiltonian for the
three and two particle systems allows for the calculation of the exact
many-body Green's function which agrees well with the GW result: it
has two poles corresponding to a singlet and a triplet state of the
two particle system separated by $U$.

\emph{NV$^-$ center}.--- Next, we apply our approach to the NV$^{-}$
center in diamond which has a triplet ground state.  This defect
complex currently attracts much attention because of its extraordinary
properties, such as long coherence times and potential application to
quantum computing \cite{Wrachtrup,Hanson,Awschalom}.

Again, we carry out DFT calculations as described in the previous
sections. We employ a 64-atom supercell and relax all
ionic positions. As a test, we first carry out GW calculations with a
generalized plasmon pole model \cite{DeslippeBGW} and find good
agreement with the similarly calculated results of Ma and Rohlfing
\cite{MaRohlfing}, who use a 256-atom supercell, for the energy
differences of the defect levels in the gap. This indicates that our
supercell size is sufficient to extract accurate multiplet splittings
for the defect levels using the current method.

For the NV$^-$ center we find that the DFT orbital energies are much
closer to the static COHSEX results than in NO$_2$, and we can use the
DFT energies and wave functions both for the construction of $W_0$ and
$G_0$. We use 300 empty orbitals and a 12 Ry momentum space cutoff for
$W_0$ and 300 empty orbitals in conjunction with a modified remainder
correction for the self-energy matrix elements. This choice of
parameters results in multiplet splittings converged to within $\sim
0.1$~eV.

\begin{figure}
  \includegraphics[width=8.cm]{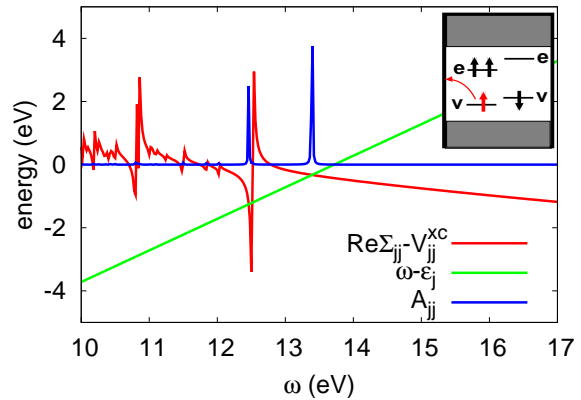}
  \caption{Self energy $\Sigma_{jj}(\omega)$ and spectral function
    $A_{jj}(\omega)$ for the removal of an up-spin electron from the
    $j=\nu$ orbital. A Lorentzian broadening of 5~meV is used for each
    curve.}
\label{fig:NVU}
\end{figure}

\begin{table}
  \setlength{\doublerulesep}{1\doublerulesep}
  \setlength{\tabcolsep}{2\tabcolsep}
  \caption{Comparison of the calculated multiplet splittings for the
    NV$^{-}$ defect in diamond with results from exact diagonalization
    calculations on the extended Hubbard model\cite{Sang}. All
    energies are given in eV.}
  \begin{ruledtabular}
    
    \begin{tabular}{c c c}
      splitting  & GW & Ref.~\cite{Sang}  \\
      \hline
      $E_{( ^2E )}-E_{(^2A_2)}$   &  2.0  &  1.8  \\
      $E_{(^4A_2)}-E_{(^2A_2)}$ &  0.9  &  0.9  \\
    \end{tabular}

  \end{ruledtabular}
  \label{NVdat}
\end{table}

Figure~\ref{fig:NVU} shows results for the removal of an
\emph{up-spin} electron from the $\nu$ level. As in the NO$_2$
calculation, the self energy exhibits a low-lying pole leading to
\emph{two} solutions of the quasiparticle equation. To understand
which many-body states these solutions correspond to, we compare our
results to exact diagonalization calculations of the extended Hubbard
model of Choi, Jain and Louie\cite{Sang}. These authors fit the
parameters of an extended Hubbard model for the defect levels to
\emph{ab initio} static COHSEX results and show that this model
describes accurately neutral excitations. The model predicts
\emph{four} many-body states $\Psi_\lambda$ for the $\nu^1e^2$
configuration. However, only \emph{two} of the four states, namely
$^4A_2$ and $^2A_2$, are observed in our calculations because by
symmetry only these states have a non-vanishing matrix element
$\langle \Psi_\lambda| c_{\nu\uparrow}| \nu_\uparrow \nu_\downarrow
e_{x\uparrow}e_{y\uparrow}\rangle$ with the ground
state. Table~\ref{NVdat} shows that we obtain good agreement with the
extended Hubbard model results for the multiplet splittings. Note that
the $^4A_2$-$^2A_2$ splitting in Fig.~\ref{fig:NVU} corresponds to the
last row in Table~\ref{NVdat}. The first row in Table~\ref{NVdat}
shows the splitting between the $^2E$ state obtained by removing an
up-spin electron from the $e$ defect orbital and the $^2A_2$ state.
Again, we find good agreement with the extended Hubbard model results.

J.L. acknowledges useful discussions with SangKook Choi and Dr. Eric
Shirley.  J.D. acknowledges support under National Science Foundation
Grant No. DMR10-1006184. J.L. and M.J. acknowledge support from the
Director, Office of Science, Office of Basic Energy Sciences,
Materials Sciences and Engineering Division, U.S. Department of Energy
under Contract No. DE- AC02-05CH11231.  Computational resources have
been provided by DOE at Lawrence Berkeley National Laboratory's NERSC
facility.

\bibliography{paper}
\end{document}